\documentclass[reprint, amsmath, amssymb, aps, prl]{revtex4-2}
\usepackage{graphicx}
\usepackage{dcolumn}
\usepackage{bm}
\usepackage{braket}
\usepackage{natbib}
\usepackage{setspace}

\begin{document}

\title{Selective observation of enantiomeric chiral phonons in $\alpha$-quartz}


\author{Eiichi Oishi}
 \email{oishi.e17@kore-lab.org}
\author{Yasuhiro Fujii}%
\author{Akitoshi Koreeda}%
\affiliation{Department of Physical Sciences, College of Science and Engineering, Ritsumeikan University, Kusatsu, Shiga 525-8577, Japan
}%
\date{\today}

\begin{abstract}
We report anomalous circularly polarized Raman spectra of phonons in right- and left-handed quartzes. The phonon branches splitting from the E-mode at a finite wavenumber were found to be chiral with mutually opposite angular momenta. Our analysis reveals the Raman selection rules for chiral phonons. We also find that the conservation of angular momentum should be satisfied between two photons and one chiral phonon in the Raman process. Furthermore, we experimentally proved that the helicity of phonons with a certain frequency and wavevector should be opposite in the other enantiomer.
\end{abstract}


\maketitle
Chiral materials have been investigated in many studies because they exhibit various properties, such as optical activity \cite{Barron2004,hecht2002optics}, circular dichroism \cite{hecht2002optics,Barron2004}, and chirality-induced spin selectivity \cite{Naaman2012,Naaman2015,Crnb2015}. In particular, chiral phonons in these materials have recently attracted significant attention \cite{Kishine2020,Chen2022,Ishito2022}. Chiral phonons are phonons with angular momentum, which correspond to lattice vibration modes with the rotation of atomic displacements \cite{Zhang2014}. Chiral phonons may contribute to the thermal Hall effect, where heat flow is deflected by a magnetic field \cite{Grissonnanche2020}. In addition, chiral phonons are expected to be associated with interesting phenomena such as coupling between chiral phonons and circularly polarized light or valley electrons \cite{Zhu2018a,Li2019,Chen2019,Chen2021}, conversion of magnons to chiral phonons \cite{Holanda2018}, phonon thermal Edelstein effects \cite{Hamada2018}, and phonon rotoelectric effects \cite{Hamada2020}. Furthermore, the degrees of freedom of angular momentum can be used for information communication technology \cite{Zhu2016}.

It is well known that the rotation direction of the plane of linear polarization owing to optical activity is opposite in right- and left-handed chiral crystals. Analogous to optical activity, chiral phonons are expected to exhibit opposite properties including angular momentum and propagation direction in crystals with different chirality. A typical example in this regard is the recently proposed chiral phonon diode effect \cite{Chen2022}. In this effect, chiral phonons along the $c$-axis of chiral crystals (e.g., $\alpha$-quartz and Te) belonging to a certain branch can propagate in only either a positive or a negative direction. In addition, it is predicted that the direction along which the chiral phonons can propagate is opposite in right- and left-handed chiral crystals. Investigation of chiral phonons in crystals with different chirality is significant for controlling angular momentum and propagation direction of chiral phonons. However, to the best of our knowledge, such experimental studies have rarely been reported till date.

Raman spectroscopy is a popular method adopted for observing phonons \cite{Loudon1964}. Particularly, Raman spectroscopy with circularly polarized light, implying photons with a spin angular momentum (SAM), is expected to be advantageous for observing the angular momentum of chiral phonons. However, the Raman selection rules of chiral phonons have not been clarified despite many reports of circularly polarized Raman spectroscopy in chiral crystals \cite{A.S.PineandG.Dresselhaus1969,Pine1971,Briggs1977,Ouillon1994,Pinan-Lucarre1999,Garasevich1995,Rodriguez1977,Bizek1981}. In this study, we aim to achieve the selective observation of enantiomeric chiral phonons in right- and left-handed quartzes by employing circularly polarized Raman spectroscopy. The experimentally deduced Raman selection rules show that phonons having a finite wavenumber can carry angular momenta and that the helicity of phonons with a certain frequency and wavevector should be the opposite in the other enantiomer.

\begin{figure}
\includegraphics{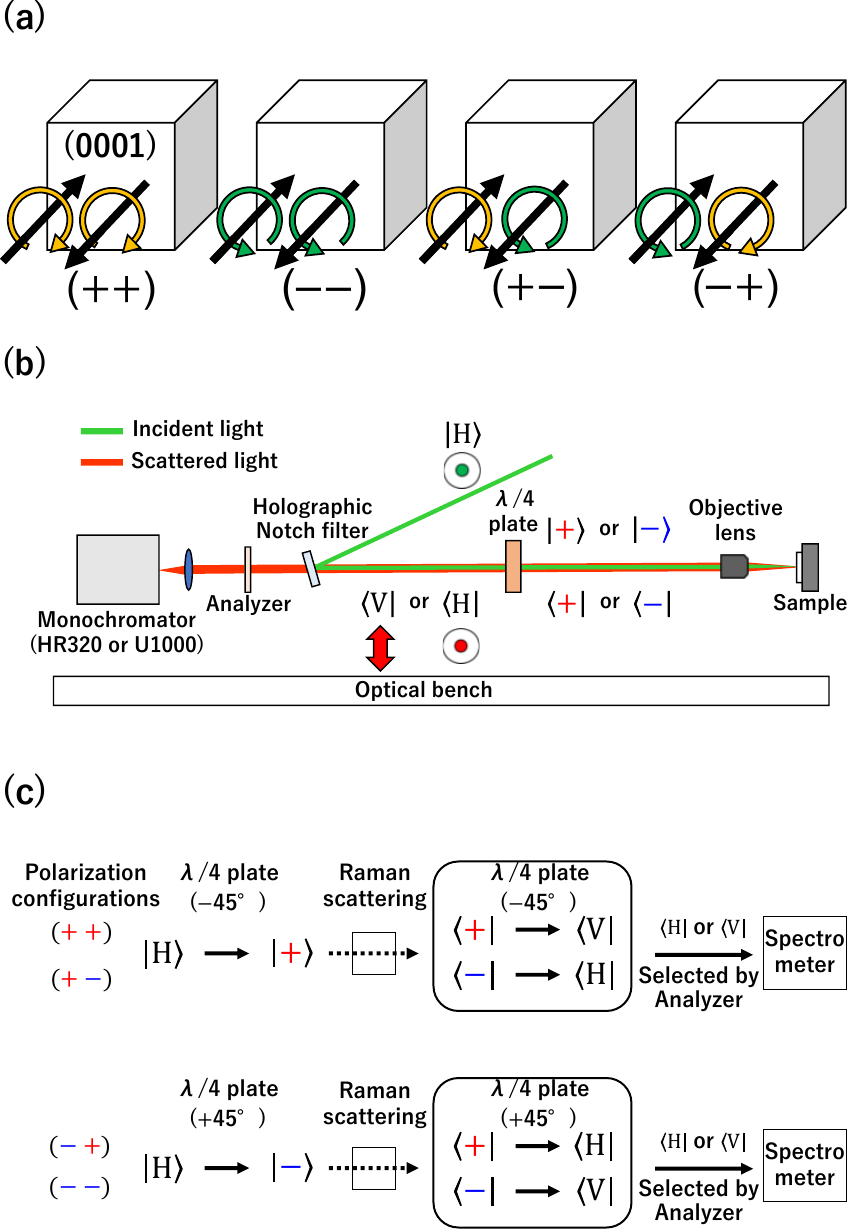}
\caption{\label{fig1} (a)Polarization configurations 
with a backscattering geometry, where $+$ and $-$ represent circularly polarized light with SAM of $+\hbar$ and $-\hbar$, respectively. (b)Raman optical system. (c)Schematic of the polarization state of light in the $(+-)$, $(-+)$, $(++)$, and $(--)$ configurations.}
\end{figure}

We employed (0001) plates of right-handed (space group P$3_1$21; Crystal Base Co., Ltd., Osaka, Japan) and left-handed quartzes (space group P$3_2$21; CRYSTAL GmbH, Berlin, Germany) with dimensions of $10\times 10\times 0.5$~mm$^3$ synthesized by hydrothermal method. We performed Raman spectroscopy on the crystals at room temperature (297~K) using a 532~nm excitation line from a frequency-doubled Nd:YAG laser (Oxxius LCX-532S-300), single monochromator (Jobin Yvon HR320 with a 1200~gr/mm grating and frequency resolution of 2.7~cm$^{-1}$), and double monochromator (Jobin Yvon U1000 with 1800~gr/mm grating and frequency resolution of 0.3~cm$^{-1}$) \cite[S1]{SM}. As shown in Fig.~\ref{fig1}(a), we adopted a backscattering geometry and four polarization configurations $(++)$, $(--)$, $(+-)$, and $(-+)$, where the first and second signs in each parenthesis indicate the sign of SAM along the incident laser direction of the incident and scattered photons, respectively. The incident optical intensity was held constant for the four polarization configurations. The SAM of the right-/left-circularly polarized light propagating along the direction toward the sample was $+\hbar/-\hbar$, respectively, while that of the backpropagating right-/left-circularly polarized light was $-\hbar/+\hbar$, respectively. In other words, the SAM signs of the photons corresponded to the rotational direction of the electric field of circularly polarized light on the sample surface regardless of the propagation direction. For example, in the polarization configuration (++), the incident light is right-circularly polarized, while the scattered light is left-circularly polarized. The SAM of the right-circularly polarized light propagating along the $+z$-direction is $+\hbar$, while that of the left-circularly polarized light propagating along the $-z$-direction is $-(-\hbar)=+\hbar$. The same concept can be used for other polarization configurations. Figures~\ref{fig1}(b) and \ref{fig1}(c) show our Raman optical system and the polarization states of lights between the laser and spectrometers in the configurations of $(+-)$, $(-+)$, $(++)$, and $(--)$, respectively. Here $\ket{\mathrm{H}}$ and $\ket{\mathrm{V}}$ indicate the lights with horizontal and vertical polarizations relative to the optical bench. We converted the linearly polarized light into circularly polarized light using a $\lambda/4$ plate, and vice versa. The circularly polarized Raman scattered light passed through the $\lambda/4$ plate and was converted into linearly polarized light. The scattered circularly polarized lights having the same and opposite signs as that of the incident light are converted to $\bra{\mathrm{V}}$ and $\bra{\mathrm{H}}$ by the $\lambda/4$ plate, respectively. Therefore, the Raman scattered lights in the $(+-)$ and $(-+)$ ($(++)$ and $(--)$) configurations were in the same linear polarization state when these lights entered the spectrometers.

\begin{figure}
\includegraphics{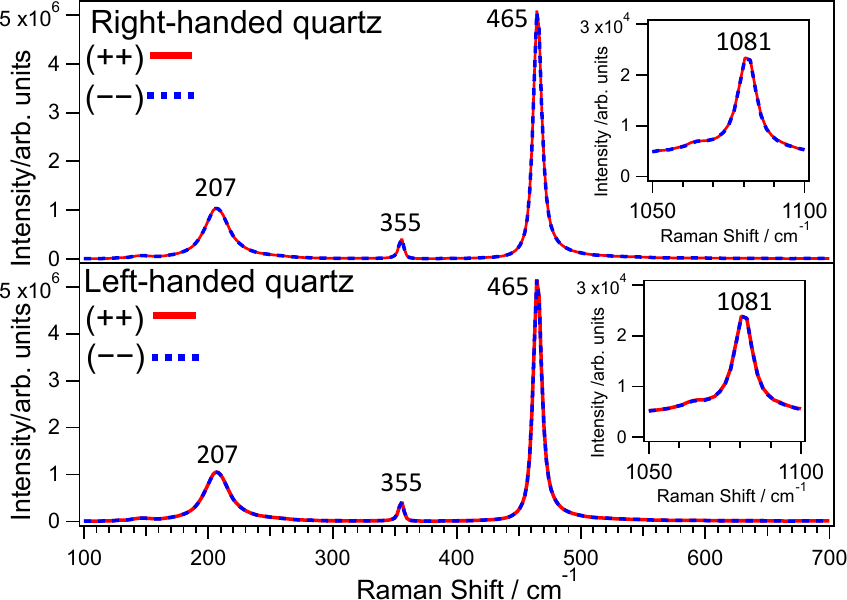}
\caption{\label{fig2} Raman spectra of right- and left-handed quartzes in polarization configurations of $(++)$ (solid red line) and $(--)$ (blue dotted line), where only A$_1$-mode is Raman active. The Raman spectra in the $(++)$ and $(--)$ configurations show a similar shape in both right- and left-handed quartzes.
}
\end{figure}

\begin{figure*}
\includegraphics{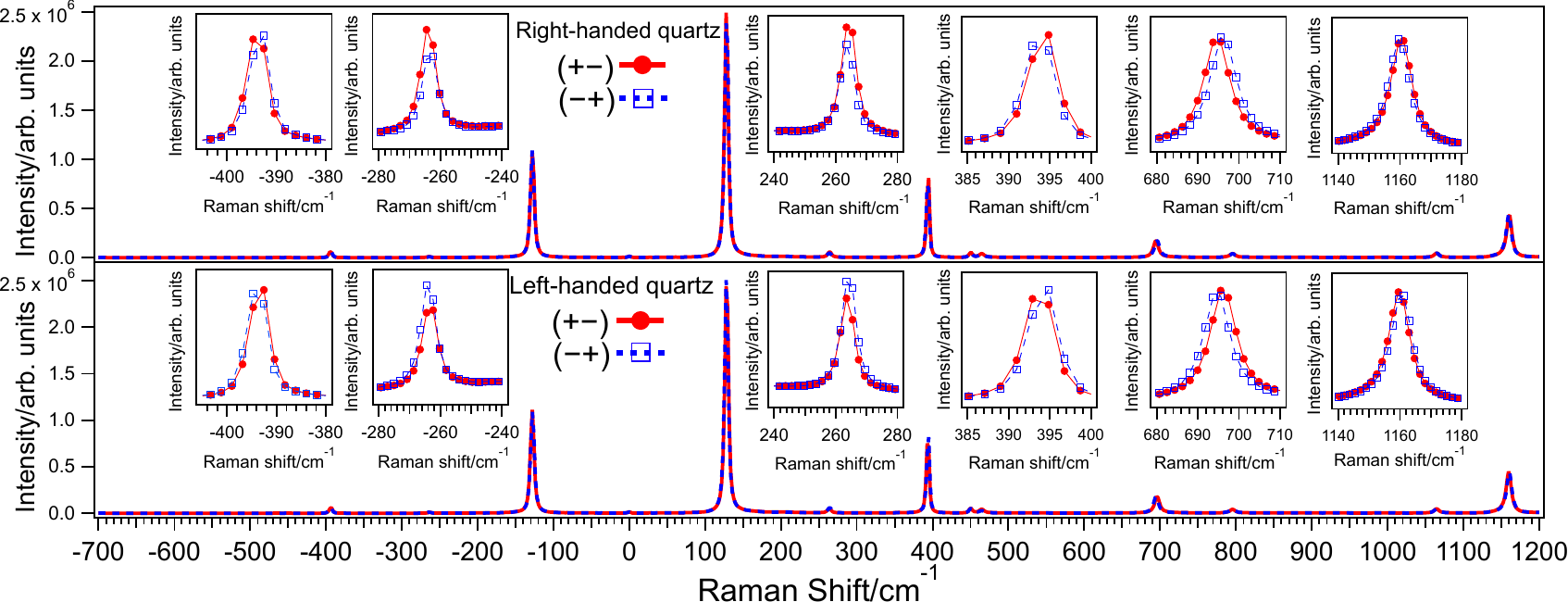}
\caption{\label{fig3} Raman spectra of right and left-handed quartzes in the polarization configurations of $(+-)$ (solid red line with circles) and $(-+)$ (dotted blue line with squares), where only E-mode is Raman active. The peak positions in the $(+-)$ and $(-+)$ configurations are slightly different for all modes. The relationship between the positions in the $(+-)$ and $(-+)$ configurations is interchanged in the right- and left-handed quartzes.}
\end{figure*}

The Raman spectra of the right- and left-handed quartzes measured with a resolution of 2.7~cm$^{-1}$ are shown in Figs.~\ref{fig2} and~\ref{fig3}, respectively. A$_1$-mode phonons were observed in $(++)$ and $(--)$ configurations, while E-mode phonons were observed in $(+-)$ and $(-+)$ configurations. Notably, each Raman shift observed in the $(+-)$ and $(-+)$ configurations was slightly different from that in the $(++)$ and $(--)$ configurations. In contrast, we confirmed that the Raman spectra of silicon (Si), which is an achiral material, in the $(+-)$ and $(-+)$ configurations showed the same Raman shift~\cite[S2]{SM}. Furthermore, the slight frequency difference was intrinsic because the Raman shifts of the Stokes (above 0 cm$^{-1}$) and anti-Stokes scattering (below 0 cm$^{-1}$) coincided for all modes. Our results are consistent with that of a previous study on $\alpha$-quartz by Pine and Dresselhaus \cite{A.S.PineandG.Dresselhaus1969}, where the lowest-frequency E-mode phonon was reported to show a slight difference depending on the polarization condition; however, the handedness of the sample crystal was not discussed. In addition, the relationship between the positions of the Raman peaks observed in the $(+-)$ and $(-+)$ configurations was interchanged in the right- and left-handed quartzes, as shown in Fig.~{\ref{fig3}}. This trend was observed for all Raman peaks of the E-mode phonons.
\begin{figure}
\includegraphics{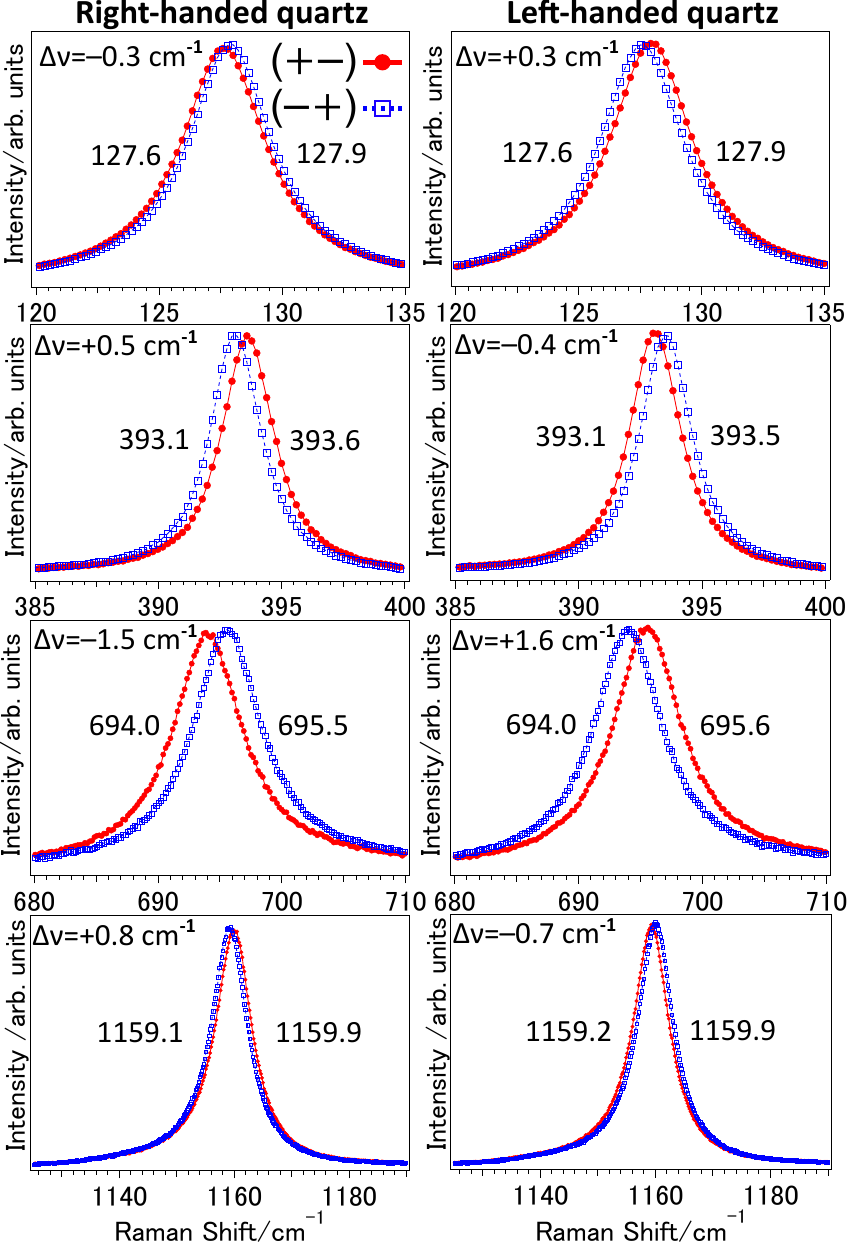}
\caption{\label{fig4} Highly resolved circularly polarized Raman spectra of the $(+-)$(solid red line with circles) and $(-+)$(dotted blue line with squares) configurations in right- and left-handed quartzes. The phonon frequency difference $\Delta\nu=\nu_{\mathrm{(+-)}}-\nu_{\mathrm{(-+)}}$ is shown in each spectrum.}
\end{figure}

To investigate the slight difference in the Raman shift, several intense Raman peaks were measured with a higher resolution of 0.3~cm$^{-1}$. The Raman peaks normalized by the integrated intensity are shown in Fig.~\ref{fig4}, where the peaks were fitted with Voigt functions to obtain the phonon frequencies $\nu_{\mathrm{(+-)}}$ and $\nu_{\mathrm{(-+)}}$. The phonon frequency difference $\Delta\nu\equiv\nu_{\mathrm{(+-)}}-\nu_{\mathrm{(-+)}}$ has been mentioned on the upper left corner of each spectrum. These results show the presence of two types of phonon modes: one is only Raman active in the $(+-)$ configuration, while the other is only Raman active in the $(-+)$ configuration. The sign of $\Delta\nu$ was found to be different for each phonon and was opposite for right- and left-handed quartzes in the same frequency region.
\begin{figure}
\includegraphics{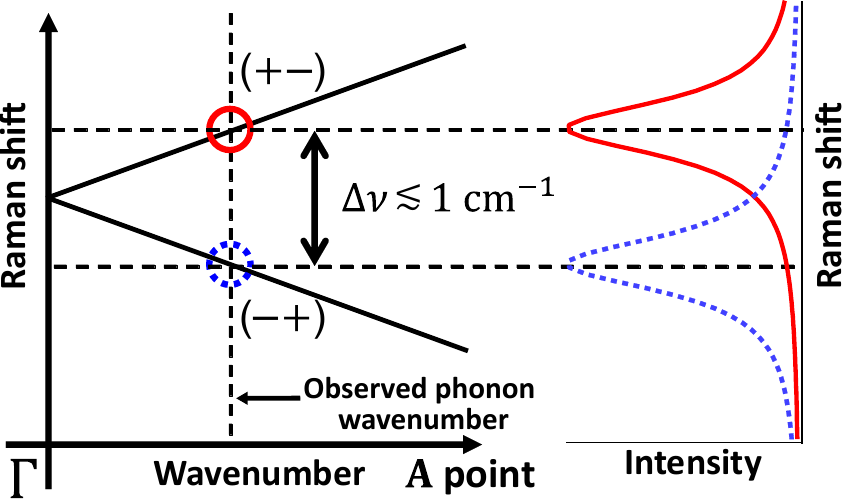}
\caption{\label{fig5} Dispersion relation for an E-mode phonon in $\alpha$-quartz. At the phonon wavenumber observed in our measurements, the E-mode phonon is split by the order of $\lesssim 1$~cm$^{-1}$, which is the same order as that of the observed $\Delta\nu$. 
}
\end{figure}

Here, we discuss the origin of $\Delta\nu$. Under the experimental conditions of this study, the observed phonons had a wavenumber of $|\bm{k}|$ from $\Gamma$ to A point, where $n$ is the refractive index of $\alpha$-quartz. According to the phonon dispersion relation of $\alpha$-quartz reported in a previous study \cite{MargaretM.Elcombe1967}, the dispersion curves of the E-mode phonons were split into two branches at a finite wavenumber. The split widths of the branches at $|\bm{k}|=4n\pi/532$~nm$^{-1}$ were $\lesssim 1$~cm$^{-1}$ \cite[S3-S5]{SM}; this agrees with the observed values of $\Delta \nu$ (Fig.~\ref{fig5}). Although circular birefringence may lead to frequency differences, it was negligible because its order was estimated to be $\lesssim 10^{-4}$~cm$^{-1}$ \cite{aB}. Thus, we consider that each pair of Raman peaks is assigned to the two phonon branches having different selection rules.
Let $\bm{H}_{\lambda}$ be the Hamiltonian of the $\lambda$-th phonon mode  expanded with respect to wavevector $\bm{q}$ as
\begin{equation}
\bm{H}_{\lambda}=\bm{H}_{\lambda 0}+\bm{q}\otimes \bm{S}_\lambda,
\end{equation}
where $\bm{S}_{\lambda}$ is the order-1 tensor of the expansion coefficients and $\bm{q}\otimes \bm{S}_\lambda$ is
\begin{equation}\label{qs_matrix}
    \bm{q}\otimes \bm{S}_\lambda = \left(
      \begin{array}{cc}
        0 & -iB_{\lambda}q_z  \\
        iB_{\lambda}q_z & 0
      \end{array}
    \right),
\end{equation}
from the crystal symmetry of $\alpha$-quartz~\cite{Rodriguez1977,Bizek1981}. Here, $B_{\lambda}$ is a real constant and $q_z$ is the $z$ component of the phonon wavenumber. In this study, we only consider a $2\times2$ matrix related to the E-mode phonon of $\bm{q}\otimes \bm{S}_\lambda$, which is essentially a $3\times3$ matrix, because it is sufficient for describing the back-scattering experiment. By introducing the eigenenergy of the unperturbed Hamiltonian $\bm{H}_{\lambda0}$ of the $\lambda$-th E-mode phonon as $E_{\lambda0}$, $\bm{H}_\lambda$ can be expressed as
\begin{equation}
  \bm{H}_\lambda = \left(
    \begin{array}{cc}
      E_{\lambda0} & -iB_{\lambda0}q_z  \\
      iB_{\lambda0}q_z & E_{\lambda0}
    \end{array}
  \right).\label{Hlambda}
\end{equation}
The eigenvectors $\left.{|u}_\lambda\right\rangle$ and eigenvalues $E_\lambda$ of $\bm{H}_\lambda$ were
\begin{equation}
  \label{eigenvalue}
E_{\lambda \pm}=E_{\lambda 0}\pm B_{\lambda}q_z,
\end{equation}
\begin{equation}
    \label{eigenvector}
\left|u_{\lambda\pm}\right\rangle=\frac{1}{\sqrt{2}}\left(
    \begin{array}{c}
      1 \\
      \pm i
    \end{array}
  \right),
\end{equation}
respectively. The subscripts of $E_{\lambda\pm}$ and $\left|u_{\lambda\pm}\right\rangle$ are in the same order as that of the double signs. Equation (\ref{eigenvalue}) expresses that the E-mode phonons split into the upper and lower branches at a finite wavenumber. Therefore, the splitting of the E-mode phonons is interpreted to have caused the observed $\Delta\nu$s. Further, the Raman spectra for the $(+-)$ and $(-+)$ configurations showed that the Raman selection rules were different for the phonons in the upper and lower branches \cite{aBC}. In addition, Eq.~(\ref{eigenvector}) implies that these phonons are in mutually opposite circular polarization states such that they are chiral phonons with opposite angular momenta.

Raman selection can be described using the Raman tensor $\bm{\alpha}$. The scattering intensity $I$ is expressed as $I=|\braket{{\bm{e}_{{\mathrm{s}}}|\bm{\alpha}|\bm{e}_{{\mathrm{i}}}}}|^2$ using the polarization vectors of incident $\bm{e}_\mathrm{i}$ and scattered $\bm{e}_\mathrm{s}$ lights. Loudon~\cite{Loudon1964} reported real-valued Raman tensors of E$(x)$- and E$(y)$-mode phonons (with normal coordinates of $x$ and $y$) for the point group $D_3$. The Raman tensors are derived by assuming linearly polarized phonons degenerated at $\Gamma$ point. However, since the phonons we observed are circularly polarized with a finite wavenumber, we need to rewrite the Raman tensors. In fact, using the Raman tensors of E$(x)$- and E$(y)$-modes, the two modes are Raman active in both $(+-)$ and $(-+)$ configurations. Therefore, we assume Raman tensors with complex elements and deduce the Raman tensors of chiral phonons from the experimental results.
Let the Raman tensor and right- and left-circularly polarized vectors be
\begin{equation}
\bm{\alpha}= \left(
  \begin{array}{cc}
    a & b  \\
  c & d
  \end{array}
\right) ,\bm{e}_+=\frac{1}{\sqrt{2}}\left(
    \begin{array}{c}
      1 \\
       -i
    \end{array}
  \right), \mathrm{and} \ \bm{e}_-=\frac{1}{\sqrt{2}}\left(
      \begin{array}{c}
        1 \\
         +i
      \end{array}
    \right) \nonumber,
\end{equation}
respectively, where $a$, $b$, $c$, and $d$ are complex constants. As an example, we derive the Raman tensor of a phonon mode whose Raman scattering intensity was non-zero only in the $(+-)$ configuration; the Raman peak (solid red line with circles) near 128 cm$^{-1}$ in right-handed quartz shown in Fig.~\ref{fig4}. The Raman scattering intensity of the phonon mode had a finite value in the $(+-)$ configuration, while it vanished in other polarization configurations. Therefore, the experimental results are expressed as $I_{\mathrm{(++)}}=I_{\mathrm{(--)}}=I_{\mathrm{(-+)}}=0$ and $I_{\mathrm{(+-)}}=A$, where $A$ is a real-valued constant. On solving, we obtain
$\bm{\alpha}_{\lambda}=C_{\lambda}\left(
  \begin{array}{cc}
    1 & i  \\
  i & -1
  \end{array}
\right) $. Similarly, the Raman tensor of the phonon mode whose Raman scattering intensity was non-zero only in the $(-+)$ configuration is obtain as $\bm{\alpha}_{\lambda^*}=D_{\lambda}\left(
  \begin{array}{cc}
    1 & -i  \\
  -i & -1
  \end{array}
\right) $. Here, $C_{\lambda}$ and $D_{\lambda}$ are constants and the matrices $\bm{\alpha}_{\lambda}$ and $\bm{\alpha}_{\lambda^*}$ are Hermitian conjugate, reflecting the fact that we observed chiral phonons with mutually opposite angular momenta in the $(+-)$ and $(-+)$ configurations.

\begin{figure}
\includegraphics[width=86mm]{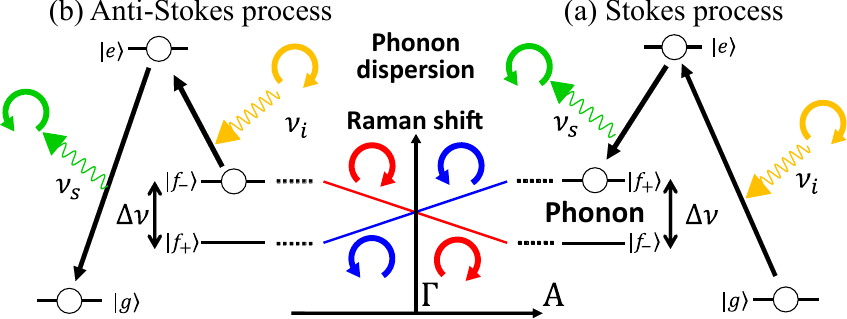}
\caption{\label{fig6} (a) (right) Stokes--Raman process and (b) (left) anti-Stokes--Raman process in the $(+-)$ configuration.
(a) Chiral phonon with a PSAM of $-\hbar$ is generated. (b) Chiral phonon with a PSAM of $+\hbar$ is annihilated.}
\end{figure}

Next, we consider the Raman scattering process of chiral phonons. In a typical Raman scattering process, energy- and momentum-conservation laws are established for two photons and one phonon \cite{Loudon1964}. In addition, the angular momentum of the photons and phonon are considered because the SAM of the incident and scattered photons differs in the $(+-)$ and $(-+)$ configurations. Accordingly, we consider the Stokes--Raman processes in the $(+-)$ and $(-+)$ configurations. The chiral phonons generated in the $(+-)$ and $(-+)$ configurations received SAMs of $+2\hbar$ and $-2\hbar$ from the incident and scattered photons, respectively. Note that the angular momentum of the phonon is a pseudo-spin angular momentum (PSAM) \cite{Zhang2015,Zhu2018a,Tatsumi2018b,Streib2021,Ishito2022}. Although the PSAM of the phonons appears to be $+2\hbar$ or $-2\hbar$, it can also be $-\hbar$ and $+\hbar$, respectively, owing to the Umklapp process of PSAM originating from the symmetry of the threefold (rotational) symmetry of $\alpha$-quartz \cite{Tatsumi2018b,Chen2022}. Therefore, chiral phonons with PSAM of $-\hbar$ and $+\hbar$ were generated for the $(+-)$ and $(-+)$ configurations, respectively. For example, the PSAMs of the E-mode phonon upper and lower branches near 694 cm$^{-1}$ with the largest $\Delta\nu$ in the right-handed quartz are $-2\hbar=+\hbar$ and $+2\hbar=-\hbar$, respectively. Our experimental results are consistent with the PSAM of phonons in right-handed quartz calculated in a recent study \cite{Chen2022}. Since the same argument can be made for the anti-Stokes--Raman process, the laws of conservation of energy, momentum, and pseudo-angular momentum for the Stokes and anti-Stokes--Raman processes can be expressed as
\begin{align}
 h\nu_\mathrm{s}-h\nu_\mathrm{i}=\mp h\nu_\mathrm{q}, \label{enegyconservation} \\
 \bm{p}_\mathrm{s}-\bm{p}_\mathrm{i}=\mp \bm{p}_\mathrm{q}, \label{momentumconservation}\\
 \bm{s}_\mathrm{s}-\bm{s}_\mathrm{i}=\mp\bm{s}_\mathrm{q}, \label{angularmomentumconservation}
\end{align}
where $\nu$, $\bm{p}$, and $\bm{s}$ are the frequency, momentum, and angular momentum, respectively, and the subscripts $\mathrm{i}$, $\mathrm{s}$, and $\mathrm{q}$ indicate the incident photon, scattered photon, and phonon, respectively. $-$ and $+$ on the right-hand sides of (\ref{enegyconservation}) -- (\ref{angularmomentumconservation}) represent the Stokes and anti-Stokes processes, respectively. The Stokes and anti-Stokes--Raman processes in the $(+-)$ configuration are shown in Fig.~\ref{fig6}, where $h$, $\nu_\mathrm{i}$, $\nu_\mathrm{s}$, $\ket{g}$, $\ket{e}$,
$\ket{f_{\mathrm{+}}}$, and $\ket{f_{\mathrm{-}}}$ are the Planck constant, frequencies of the incident and scattered lights, initial state, intermediate state, right- and left-circularly polarized states of the phonon, respectively. In the $(+-)$ Stokes--Raman process, a chiral phonon with PSAM of $-\hbar$ is generated, as shown in Fig. \ref{fig6}(a). Contrarily, a chiral phonon with PSAM of $+\hbar$ is annihilated in the $(+-)$ anti-Stokes--Raman process, as shown in Fig. \ref{fig6}(b).

Finally, we discuss the chiral phonons in right- and left-handed quartzes. As shown in Fig.~\ref{fig4}, the signs of $\Delta\nu$ for right- and left-handed quartzes were opposite to each other in the same frequency range. This suggests an interchange of the Raman symmetries of the paired chiral phonons based on the chirality of $\alpha$-quartz. Specifically, the helicities of the paired chiral phonons are switched in the other enantiomer.

To summarize, we selectively observed enantiomeric chiral phonons in right- and left-handed quartzes by circularly polarized Raman spectroscopy. The Raman selection rules of the chiral phonons were explained by complex Raman tensors
$\bm{\alpha}_{\lambda}=C_{\lambda}\left(
  \begin{array}{cc}
    1 & i  \\
  i & -1
  \end{array}
\right) $
 and
 $\bm{\alpha}_{\lambda^*}=D_{\lambda}\left(
  \begin{array}{cc}
    1 & -i  \\
  -i & -1
  \end{array}
\right) $.
We found that the law of conservation of angular momentum, in addition to those of energy and momentum, were established among the photons and chiral phonons in the Raman scattering process. Furthermore, our experimental results implied that the helicity of phonons with a certain frequency and wavevector should be opposite in the other enantiomer. In the future, the Raman tensor should be derived from a Raman cross-section considering phonons with a finite wavenumber and conservation of angular momentum. Further study is required to obtain more information on the chirality of elementary excitations.

The authors would like to thank Professor Takuya Satoh for helpful discussions on the pseudo-angular momentum. This work was supported by JST SPRING Grant Number JPMJSP2101 and JSPS KAKENHI Grant Numbers JP19K05252, JP19H05618, and JP21H01018.
\bibliography{Quartz_paper_ref.bib}

\begin{thebibliography}{34}%
\makeatletter
\providecommand \@ifxundefined [1]{%
 \@ifx{#1\undefined}
}%
\providecommand \@ifnum [1]{%
 \ifnum #1\expandafter \@firstoftwo
 \else \expandafter \@secondoftwo
 \fi
}%
\providecommand \@ifx [1]{%
 \ifx #1\expandafter \@firstoftwo
 \else \expandafter \@secondoftwo
 \fi
}%
\providecommand \natexlab [1]{#1}%
\providecommand \enquote  [1]{``#1''}%
\providecommand \bibnamefont  [1]{#1}%
\providecommand \bibfnamefont [1]{#1}%
\providecommand \citenamefont [1]{#1}%
\providecommand \href@noop [0]{\@secondoftwo}%
\providecommand \href [0]{\begingroup \@sanitize@url \@href}%
\providecommand \@href[1]{\@@startlink{#1}\@@href}%
\providecommand \@@href[1]{\endgroup#1\@@endlink}%
\providecommand \@sanitize@url [0]{\catcode `\\12\catcode `\$12\catcode
  `\&12\catcode `\#12\catcode `\^12\catcode `\_12\catcode `\%12\relax}%
\providecommand \@@startlink[1]{}%
\providecommand \@@endlink[0]{}%
\providecommand \url  [0]{\begingroup\@sanitize@url \@url }%
\providecommand \@url [1]{\endgroup\@href {#1}{\urlprefix }}%
\providecommand \urlprefix  [0]{URL }%
\providecommand \Eprint [0]{\href }%
\providecommand \doibase [0]{https://doi.org/}%
\providecommand \selectlanguage [0]{\@gobble}%
\providecommand \bibinfo  [0]{\@secondoftwo}%
\providecommand \bibfield  [0]{\@secondoftwo}%
\providecommand \translation [1]{[#1]}%
\providecommand \BibitemOpen [0]{}%
\providecommand \bibitemStop [0]{}%
\providecommand \bibitemNoStop [0]{.\EOS\space}%
\providecommand \EOS [0]{\spacefactor3000\relax}%
\providecommand \BibitemShut  [1]{\csname bibitem#1\endcsname}%
\let\auto@bib@innerbib\@empty
\bibitem [{\citenamefont {Barron}(2004)}]{Barron2004}%
  \BibitemOpen
  \bibfield  {author} {\bibinfo {author} {\bibfnamefont {L.~D.}\ \bibnamefont
  {Barron}},\ }\href {https://doi.org/10.1017/CBO9780511535468} {\emph
  {\bibinfo {title} {{Molecular Light Scattering and Optical Activity}}}}\
  (\bibinfo  {publisher} {Cambridge University Press},\ \bibinfo {year}
  {2004})\BibitemShut {NoStop}%
\bibitem [{\citenamefont {Hecht}(2002)}]{hecht2002optics}%
  \BibitemOpen
  \bibfield  {author} {\bibinfo {author} {\bibfnamefont {E.}~\bibnamefont
  {Hecht}},\ }\href {https://books.google.co.jp/books?id=7aG6QgAACAAJ} {\emph
  {\bibinfo {title} {Optics}}}\ (\bibinfo  {publisher} {Addison-Wesley},\
  \bibinfo {year} {2002})\BibitemShut {NoStop}%
\bibitem [{\citenamefont {Naaman}\ and\ \citenamefont
  {Waldeck}(2012)}]{Naaman2012}%
  \BibitemOpen
  \bibfield  {author} {\bibinfo {author} {\bibfnamefont {R.}~\bibnamefont
  {Naaman}}\ and\ \bibinfo {author} {\bibfnamefont {D.~H.}\ \bibnamefont
  {Waldeck}},\ }\href {https://doi.org/10.1021/jz300793y} {\bibfield  {journal}
  {\bibinfo  {journal} {J. Phys. Chem. Lett.}\ }\textbf {\bibinfo {volume}
  {3}},\ \bibinfo {pages} {2178} (\bibinfo {year} {2012})}\BibitemShut
  {NoStop}%
\bibitem [{\citenamefont {Naaman}\ and\ \citenamefont
  {Waldeck}(2015)}]{Naaman2015}%
  \BibitemOpen
  \bibfield  {author} {\bibinfo {author} {\bibfnamefont {R.}~\bibnamefont
  {Naaman}}\ and\ \bibinfo {author} {\bibfnamefont {D.~H.}\ \bibnamefont
  {Waldeck}},\ }\href {https://doi.org/10.1146/annurev-physchem-040214-121554}
  {\bibfield  {journal} {\bibinfo  {journal} {Annu. Rev. Phys. Chem.}\ }\textbf
  {\bibinfo {volume} {66}},\ \bibinfo {pages} {263} (\bibinfo {year}
  {2015})}\BibitemShut {NoStop}%
\bibitem [{\citenamefont {Inui}\ \emph {et~al.}(2020)\citenamefont {Inui},
  \citenamefont {Aoki}, \citenamefont {Nishiue}, \citenamefont {Shiota},
  \citenamefont {Kousaka}, \citenamefont {Shishido}, \citenamefont {Hirobe},
  \citenamefont {Suda}, \citenamefont {Ohe}, \citenamefont {Kishine},
  \citenamefont {Yamamoto},\ and\ \citenamefont {Togawa}}]{Crnb2015}%
  \BibitemOpen
  \bibfield  {author} {\bibinfo {author} {\bibfnamefont {A.}~\bibnamefont
  {Inui}}, \bibinfo {author} {\bibfnamefont {R.}~\bibnamefont {Aoki}}, \bibinfo
  {author} {\bibfnamefont {Y.}~\bibnamefont {Nishiue}}, \bibinfo {author}
  {\bibfnamefont {K.}~\bibnamefont {Shiota}}, \bibinfo {author} {\bibfnamefont
  {Y.}~\bibnamefont {Kousaka}}, \bibinfo {author} {\bibfnamefont
  {H.}~\bibnamefont {Shishido}}, \bibinfo {author} {\bibfnamefont
  {D.}~\bibnamefont {Hirobe}}, \bibinfo {author} {\bibfnamefont
  {M.}~\bibnamefont {Suda}}, \bibinfo {author} {\bibfnamefont {J.~I.}\
  \bibnamefont {Ohe}}, \bibinfo {author} {\bibfnamefont {J.~I.}\ \bibnamefont
  {Kishine}}, \bibinfo {author} {\bibfnamefont {H.~M.}\ \bibnamefont
  {Yamamoto}},\ and\ \bibinfo {author} {\bibfnamefont {Y.}~\bibnamefont
  {Togawa}},\ }\href {https://doi.org/10.1103/PhysRevLett.124.166602}
  {\bibfield  {journal} {\bibinfo  {journal} {Phys. Rev. Lett.}\ }\textbf
  {\bibinfo {volume} {124}},\ \bibinfo {pages} {166602} (\bibinfo {year}
  {2020})}\BibitemShut {NoStop}%
\bibitem [{\citenamefont {Kishine}\ \emph {et~al.}(2020)\citenamefont
  {Kishine}, \citenamefont {Ovchinnikov},\ and\ \citenamefont
  {Tereshchenko}}]{Kishine2020}%
  \BibitemOpen
  \bibfield  {author} {\bibinfo {author} {\bibfnamefont {J.}~\bibnamefont
  {Kishine}}, \bibinfo {author} {\bibfnamefont {A.~S.}\ \bibnamefont
  {Ovchinnikov}},\ and\ \bibinfo {author} {\bibfnamefont {A.~A.}\ \bibnamefont
  {Tereshchenko}},\ }\href {https://doi.org/10.1103/PhysRevLett.125.245302}
  {\bibfield  {journal} {\bibinfo  {journal} {Phys. Rev. Lett.}\ }\textbf
  {\bibinfo {volume} {125}},\ \bibinfo {pages} {245302} (\bibinfo {year}
  {2020})},\ \Eprint {https://arxiv.org/abs/2007.07782} {arXiv:2007.07782}
  \BibitemShut {NoStop}%
\bibitem [{\citenamefont {Chen}\ \emph {et~al.}(2022)\citenamefont {Chen},
  \citenamefont {Wu}, \citenamefont {Zhu}, \citenamefont {Yang}, \citenamefont
  {Gong}, \citenamefont {Gao}, \citenamefont {Yang},\ and\ \citenamefont
  {Zhang}}]{Chen2022}%
  \BibitemOpen
  \bibfield  {author} {\bibinfo {author} {\bibfnamefont {H.}~\bibnamefont
  {Chen}}, \bibinfo {author} {\bibfnamefont {W.}~\bibnamefont {Wu}}, \bibinfo
  {author} {\bibfnamefont {J.}~\bibnamefont {Zhu}}, \bibinfo {author}
  {\bibfnamefont {Z.}~\bibnamefont {Yang}}, \bibinfo {author} {\bibfnamefont
  {W.}~\bibnamefont {Gong}}, \bibinfo {author} {\bibfnamefont {W.}~\bibnamefont
  {Gao}}, \bibinfo {author} {\bibfnamefont {S.~A.}\ \bibnamefont {Yang}},\ and\
  \bibinfo {author} {\bibfnamefont {L.}~\bibnamefont {Zhang}},\ }\href
  {https://doi.org/10.1021/acs.nanolett.1c04705} {\bibfield  {journal}
  {\bibinfo  {journal} {Nano Lett.}\ }\textbf {\bibinfo {volume} {22}},\
  \bibinfo {pages} {1688} (\bibinfo {year} {2022})}\BibitemShut {NoStop}%
\bibitem [{\citenamefont {Ishito}\ \emph {et~al.}(2023)\citenamefont {Ishito},
  \citenamefont {Mao}, \citenamefont {Kousaka}, \citenamefont {Togawa},
  \citenamefont {Iwasaki}, \citenamefont {Zhang}, \citenamefont {Murakami},
  \citenamefont {Kishine},\ and\ \citenamefont {Satoh}}]{Ishito2022}%
  \BibitemOpen
  \bibfield  {author} {\bibinfo {author} {\bibfnamefont {K.}~\bibnamefont
  {Ishito}}, \bibinfo {author} {\bibfnamefont {H.}~\bibnamefont {Mao}},
  \bibinfo {author} {\bibfnamefont {Y.}~\bibnamefont {Kousaka}}, \bibinfo
  {author} {\bibfnamefont {Y.}~\bibnamefont {Togawa}}, \bibinfo {author}
  {\bibfnamefont {S.}~\bibnamefont {Iwasaki}}, \bibinfo {author} {\bibfnamefont
  {T.}~\bibnamefont {Zhang}}, \bibinfo {author} {\bibfnamefont
  {S.}~\bibnamefont {Murakami}}, \bibinfo {author} {\bibfnamefont {J.-i.}\
  \bibnamefont {Kishine}},\ and\ \bibinfo {author} {\bibfnamefont
  {T.}~\bibnamefont {Satoh}},\ }\href
  {https://doi.org/10.1038/s41567-022-01790-x} {\bibfield  {journal} {\bibinfo
  {journal} {Nat. Phys.}\ }\textbf {\bibinfo {volume} {19}},\ \bibinfo {pages}
  {35} (\bibinfo {year} {2023})}\BibitemShut {NoStop}%
\bibitem [{\citenamefont {Zhang}\ and\ \citenamefont {Niu}(2014)}]{Zhang2014}%
  \BibitemOpen
  \bibfield  {author} {\bibinfo {author} {\bibfnamefont {L.}~\bibnamefont
  {Zhang}}\ and\ \bibinfo {author} {\bibfnamefont {Q.}~\bibnamefont {Niu}},\
  }\href {https://doi.org/10.1103/PhysRevLett.112.085503} {\bibfield  {journal}
  {\bibinfo  {journal} {Phys. Rev. Lett.}\ }\textbf {\bibinfo {volume} {112}},\
  \bibinfo {pages} {085503} (\bibinfo {year} {2014})}\BibitemShut {NoStop}%
\bibitem [{\citenamefont {Grissonnanche}\ \emph {et~al.}(2020)\citenamefont
  {Grissonnanche}, \citenamefont {Th{\'{e}}riault}, \citenamefont {Gourgout},
  \citenamefont {Boulanger}, \citenamefont {Lefran{\c{c}}ois}, \citenamefont
  {Ataei}, \citenamefont {Lalibert{\'{e}}}, \citenamefont {Dion}, \citenamefont
  {Zhou}, \citenamefont {Pyon}, \citenamefont {Takayama}, \citenamefont
  {Takagi}, \citenamefont {Doiron-Leyraud},\ and\ \citenamefont
  {Taillefer}}]{Grissonnanche2020}%
  \BibitemOpen
  \bibfield  {author} {\bibinfo {author} {\bibfnamefont {G.}~\bibnamefont
  {Grissonnanche}}, \bibinfo {author} {\bibfnamefont {S.}~\bibnamefont
  {Th{\'{e}}riault}}, \bibinfo {author} {\bibfnamefont {A.}~\bibnamefont
  {Gourgout}}, \bibinfo {author} {\bibfnamefont {M.-E.}\ \bibnamefont
  {Boulanger}}, \bibinfo {author} {\bibfnamefont {E.}~\bibnamefont
  {Lefran{\c{c}}ois}}, \bibinfo {author} {\bibfnamefont {A.}~\bibnamefont
  {Ataei}}, \bibinfo {author} {\bibfnamefont {F.}~\bibnamefont
  {Lalibert{\'{e}}}}, \bibinfo {author} {\bibfnamefont {M.}~\bibnamefont
  {Dion}}, \bibinfo {author} {\bibfnamefont {J.-S.}\ \bibnamefont {Zhou}},
  \bibinfo {author} {\bibfnamefont {S.}~\bibnamefont {Pyon}}, \bibinfo {author}
  {\bibfnamefont {T.}~\bibnamefont {Takayama}}, \bibinfo {author}
  {\bibfnamefont {H.}~\bibnamefont {Takagi}}, \bibinfo {author} {\bibfnamefont
  {N.}~\bibnamefont {Doiron-Leyraud}},\ and\ \bibinfo {author} {\bibfnamefont
  {L.}~\bibnamefont {Taillefer}},\ }\href
  {https://doi.org/10.1038/s41567-020-0965-y} {\bibfield  {journal} {\bibinfo
  {journal} {Nat. Phys.}\ }\textbf {\bibinfo {volume} {16}},\ \bibinfo {pages}
  {1108} (\bibinfo {year} {2020})}\BibitemShut {NoStop}%
\bibitem [{\citenamefont {Zhu}\ \emph {et~al.}(2018)\citenamefont {Zhu},
  \citenamefont {Yi}, \citenamefont {Li}, \citenamefont {Xiao}, \citenamefont
  {Zhang}, \citenamefont {Yang}, \citenamefont {Kaindl}, \citenamefont {Li},
  \citenamefont {Wang},\ and\ \citenamefont {Zhang}}]{Zhu2018a}%
  \BibitemOpen
  \bibfield  {author} {\bibinfo {author} {\bibfnamefont {H.}~\bibnamefont
  {Zhu}}, \bibinfo {author} {\bibfnamefont {J.}~\bibnamefont {Yi}}, \bibinfo
  {author} {\bibfnamefont {M.-Y.}\ \bibnamefont {Li}}, \bibinfo {author}
  {\bibfnamefont {J.}~\bibnamefont {Xiao}}, \bibinfo {author} {\bibfnamefont
  {L.}~\bibnamefont {Zhang}}, \bibinfo {author} {\bibfnamefont {C.-W.}\
  \bibnamefont {Yang}}, \bibinfo {author} {\bibfnamefont {R.~A.}\ \bibnamefont
  {Kaindl}}, \bibinfo {author} {\bibfnamefont {L.-J.}\ \bibnamefont {Li}},
  \bibinfo {author} {\bibfnamefont {Y.}~\bibnamefont {Wang}},\ and\ \bibinfo
  {author} {\bibfnamefont {X.}~\bibnamefont {Zhang}},\ }\href
  {https://doi.org/10.1126/science.aar2711} {\bibfield  {journal} {\bibinfo
  {journal} {Science}\ }\textbf {\bibinfo {volume} {359}},\ \bibinfo {pages}
  {579} (\bibinfo {year} {2018})}\BibitemShut {NoStop}%
\bibitem [{\citenamefont {Li}\ \emph {et~al.}(2019)\citenamefont {Li},
  \citenamefont {Wang}, \citenamefont {Jin}, \citenamefont {Lu}, \citenamefont
  {Lian}, \citenamefont {Meng}, \citenamefont {Blei}, \citenamefont {Gao},
  \citenamefont {Taniguchi}, \citenamefont {Watanabe}, \citenamefont {Ren},
  \citenamefont {Cao}, \citenamefont {Tongay}, \citenamefont {Smirnov},
  \citenamefont {Zhang},\ and\ \citenamefont {Shi}}]{Li2019}%
  \BibitemOpen
  \bibfield  {author} {\bibinfo {author} {\bibfnamefont {Z.}~\bibnamefont
  {Li}}, \bibinfo {author} {\bibfnamefont {T.}~\bibnamefont {Wang}}, \bibinfo
  {author} {\bibfnamefont {C.}~\bibnamefont {Jin}}, \bibinfo {author}
  {\bibfnamefont {Z.}~\bibnamefont {Lu}}, \bibinfo {author} {\bibfnamefont
  {Z.}~\bibnamefont {Lian}}, \bibinfo {author} {\bibfnamefont {Y.}~\bibnamefont
  {Meng}}, \bibinfo {author} {\bibfnamefont {M.}~\bibnamefont {Blei}}, \bibinfo
  {author} {\bibfnamefont {M.}~\bibnamefont {Gao}}, \bibinfo {author}
  {\bibfnamefont {T.}~\bibnamefont {Taniguchi}}, \bibinfo {author}
  {\bibfnamefont {K.}~\bibnamefont {Watanabe}}, \bibinfo {author}
  {\bibfnamefont {T.}~\bibnamefont {Ren}}, \bibinfo {author} {\bibfnamefont
  {T.}~\bibnamefont {Cao}}, \bibinfo {author} {\bibfnamefont {S.}~\bibnamefont
  {Tongay}}, \bibinfo {author} {\bibfnamefont {D.}~\bibnamefont {Smirnov}},
  \bibinfo {author} {\bibfnamefont {L.}~\bibnamefont {Zhang}},\ and\ \bibinfo
  {author} {\bibfnamefont {S.-F.}\ \bibnamefont {Shi}},\ }\href
  {https://doi.org/10.1021/acsnano.9b06682} {\bibfield  {journal} {\bibinfo
  {journal} {ACS Nano}\ }\textbf {\bibinfo {volume} {13}},\ \bibinfo {pages}
  {14107} (\bibinfo {year} {2019})}\BibitemShut {NoStop}%
\bibitem [{\citenamefont {Chen}\ \emph {et~al.}(2019)\citenamefont {Chen},
  \citenamefont {Lu}, \citenamefont {Dubey}, \citenamefont {Yao}, \citenamefont
  {Liu}, \citenamefont {Wang}, \citenamefont {Xiong}, \citenamefont {Zhang},\
  and\ \citenamefont {Srivastava}}]{Chen2019}%
  \BibitemOpen
  \bibfield  {author} {\bibinfo {author} {\bibfnamefont {X.}~\bibnamefont
  {Chen}}, \bibinfo {author} {\bibfnamefont {X.}~\bibnamefont {Lu}}, \bibinfo
  {author} {\bibfnamefont {S.}~\bibnamefont {Dubey}}, \bibinfo {author}
  {\bibfnamefont {Q.}~\bibnamefont {Yao}}, \bibinfo {author} {\bibfnamefont
  {S.}~\bibnamefont {Liu}}, \bibinfo {author} {\bibfnamefont {X.}~\bibnamefont
  {Wang}}, \bibinfo {author} {\bibfnamefont {Q.}~\bibnamefont {Xiong}},
  \bibinfo {author} {\bibfnamefont {L.}~\bibnamefont {Zhang}},\ and\ \bibinfo
  {author} {\bibfnamefont {A.}~\bibnamefont {Srivastava}},\ }\href
  {https://doi.org/10.1038/s41567-018-0366-7} {\bibfield  {journal} {\bibinfo
  {journal} {Nat. Phys.}\ }\textbf {\bibinfo {volume} {15}},\ \bibinfo {pages}
  {221} (\bibinfo {year} {2019})},\ \Eprint {https://arxiv.org/abs/1808.09984}
  {arXiv:1808.09984} \BibitemShut {NoStop}%
\bibitem [{\citenamefont {Chen}\ \emph {et~al.}(2021)\citenamefont {Chen},
  \citenamefont {Wu}, \citenamefont {Zhu}, \citenamefont {Yang},\ and\
  \citenamefont {Zhang}}]{Chen2021}%
  \BibitemOpen
  \bibfield  {author} {\bibinfo {author} {\bibfnamefont {H.}~\bibnamefont
  {Chen}}, \bibinfo {author} {\bibfnamefont {W.}~\bibnamefont {Wu}}, \bibinfo
  {author} {\bibfnamefont {J.}~\bibnamefont {Zhu}}, \bibinfo {author}
  {\bibfnamefont {S.~A.}\ \bibnamefont {Yang}},\ and\ \bibinfo {author}
  {\bibfnamefont {L.}~\bibnamefont {Zhang}},\ }\href
  {https://doi.org/10.1021/acs.nanolett.1c00236} {\bibfield  {journal}
  {\bibinfo  {journal} {Nano Lett.}\ }\textbf {\bibinfo {volume} {21}},\
  \bibinfo {pages} {3060} (\bibinfo {year} {2021})}\BibitemShut {NoStop}%
\bibitem [{\citenamefont {Holanda}\ \emph {et~al.}(2018)\citenamefont
  {Holanda}, \citenamefont {Maior}, \citenamefont {Azevedo},\ and\
  \citenamefont {Rezende}}]{Holanda2018}%
  \BibitemOpen
  \bibfield  {author} {\bibinfo {author} {\bibfnamefont {J.}~\bibnamefont
  {Holanda}}, \bibinfo {author} {\bibfnamefont {D.~S.}\ \bibnamefont {Maior}},
  \bibinfo {author} {\bibfnamefont {A.}~\bibnamefont {Azevedo}},\ and\ \bibinfo
  {author} {\bibfnamefont {S.~M.}\ \bibnamefont {Rezende}},\ }\href
  {https://doi.org/10.1038/s41567-018-0079-y} {\bibfield  {journal} {\bibinfo
  {journal} {Nat. Phys.}\ }\textbf {\bibinfo {volume} {14}},\ \bibinfo {pages}
  {500} (\bibinfo {year} {2018})}\BibitemShut {NoStop}%
\bibitem [{\citenamefont {Hamada}\ \emph {et~al.}(2018)\citenamefont {Hamada},
  \citenamefont {Minamitani}, \citenamefont {Hirayama},\ and\ \citenamefont
  {Murakami}}]{Hamada2018}%
  \BibitemOpen
  \bibfield  {author} {\bibinfo {author} {\bibfnamefont {M.}~\bibnamefont
  {Hamada}}, \bibinfo {author} {\bibfnamefont {E.}~\bibnamefont {Minamitani}},
  \bibinfo {author} {\bibfnamefont {M.}~\bibnamefont {Hirayama}},\ and\
  \bibinfo {author} {\bibfnamefont {S.}~\bibnamefont {Murakami}},\ }\href
  {https://doi.org/10.1103/PhysRevLett.121.175301} {\bibfield  {journal}
  {\bibinfo  {journal} {Phys. Rev. Lett.}\ }\textbf {\bibinfo {volume} {121}},\
  \bibinfo {pages} {175301} (\bibinfo {year} {2018})}\BibitemShut {NoStop}%
\bibitem [{\citenamefont {Hamada}\ and\ \citenamefont
  {Murakami}(2020)}]{Hamada2020}%
  \BibitemOpen
  \bibfield  {author} {\bibinfo {author} {\bibfnamefont {M.}~\bibnamefont
  {Hamada}}\ and\ \bibinfo {author} {\bibfnamefont {S.}~\bibnamefont
  {Murakami}},\ }\href {https://doi.org/10.1103/PhysRevB.101.144306} {\bibfield
   {journal} {\bibinfo  {journal} {Phys. Rev. B}\ }\textbf {\bibinfo {volume}
  {101}},\ \bibinfo {pages} {144306} (\bibinfo {year} {2020})}\BibitemShut
  {NoStop}%
\bibitem [{\citenamefont {Zhu}\ \emph {et~al.}(2016)\citenamefont {Zhu},
  \citenamefont {Gao}, \citenamefont {Mu},\ and\ \citenamefont {Li}}]{Zhu2016}%
  \BibitemOpen
  \bibfield  {author} {\bibinfo {author} {\bibfnamefont {Z.}~\bibnamefont
  {Zhu}}, \bibinfo {author} {\bibfnamefont {W.}~\bibnamefont {Gao}}, \bibinfo
  {author} {\bibfnamefont {C.}~\bibnamefont {Mu}},\ and\ \bibinfo {author}
  {\bibfnamefont {H.}~\bibnamefont {Li}},\ }\href
  {https://doi.org/10.1364/optica.3.000212} {\bibfield  {journal} {\bibinfo
  {journal} {Optica}\ }\textbf {\bibinfo {volume} {3}},\ \bibinfo {pages} {212}
  (\bibinfo {year} {2016})}\BibitemShut {NoStop}%
\bibitem [{\citenamefont {Loudon}(1964)}]{Loudon1964}%
  \BibitemOpen
  \bibfield  {author} {\bibinfo {author} {\bibfnamefont {R.}~\bibnamefont
  {Loudon}},\ }\href {https://doi.org/10.1080/00018736400101051} {\bibfield
  {journal} {\bibinfo  {journal} {Adv. Phys.}\ }\textbf {\bibinfo {volume}
  {13}},\ \bibinfo {pages} {423} (\bibinfo {year} {1964})}\BibitemShut
  {NoStop}%
\bibitem [{\citenamefont {{A. S. Pine and G.
  Dresselhaus}}(1969)}]{A.S.PineandG.Dresselhaus1969}%
  \BibitemOpen
  \bibfield  {author} {\bibinfo {author} {\bibnamefont {{A. S. Pine and G.
  Dresselhaus}}},\ }\href@noop {} {\bibfield  {journal} {\bibinfo  {journal}
  {Phys. Rev.}\ }\textbf {\bibinfo {volume} {188}},\ \bibinfo {pages} {1489}
  (\bibinfo {year} {1969})}\BibitemShut {NoStop}%
\bibitem [{\citenamefont {Pine}\ and\ \citenamefont
  {Dresselhaus}(1971)}]{Pine1971}%
  \BibitemOpen
  \bibfield  {author} {\bibinfo {author} {\bibfnamefont {A.~S.}\ \bibnamefont
  {Pine}}\ and\ \bibinfo {author} {\bibfnamefont {G.}~\bibnamefont
  {Dresselhaus}},\ }\href {https://doi.org/10.1103/PhysRevB.4.356} {\bibfield
  {journal} {\bibinfo  {journal} {Phys. Rev. B}\ }\textbf {\bibinfo {volume}
  {4}},\ \bibinfo {pages} {356} (\bibinfo {year} {1971})}\BibitemShut {NoStop}%
\bibitem [{\citenamefont {Briggs}\ and\ \citenamefont
  {Ramdas}(1977)}]{Briggs1977}%
  \BibitemOpen
  \bibfield  {author} {\bibinfo {author} {\bibfnamefont {R.~J.}\ \bibnamefont
  {Briggs}}\ and\ \bibinfo {author} {\bibfnamefont {A.~K.}\ \bibnamefont
  {Ramdas}},\ }\href {https://doi.org/10.1103/PhysRevB.16.3815} {\bibfield
  {journal} {\bibinfo  {journal} {Phys. Rev. B}\ }\textbf {\bibinfo {volume}
  {16}},\ \bibinfo {pages} {3815} (\bibinfo {year} {1977})}\BibitemShut
  {NoStop}%
\bibitem [{\citenamefont {Ouillon}\ \emph {et~al.}(1994)\citenamefont
  {Ouillon}, \citenamefont {Lucarre},\ and\ \citenamefont
  {Ranson}}]{Ouillon1994}%
  \BibitemOpen
  \bibfield  {author} {\bibinfo {author} {\bibfnamefont {R.}~\bibnamefont
  {Ouillon}}, \bibinfo {author} {\bibfnamefont {J.~P.~P.}\ \bibnamefont
  {Lucarre}},\ and\ \bibinfo {author} {\bibfnamefont {P.}~\bibnamefont
  {Ranson}},\ }\href@noop {} {\bibfield  {journal} {\bibinfo  {journal} {J.
  RAMAN Spectrosc.}\ }\textbf {\bibinfo {volume} {25}},\ \bibinfo {pages} {489}
  (\bibinfo {year} {1994})}\BibitemShut {NoStop}%
\bibitem [{\citenamefont {Pinan-Lucarre}\ \emph {et~al.}(1999)\citenamefont
  {Pinan-Lucarre}, \citenamefont {Ouillon},\ and\ \citenamefont
  {Ranson}}]{Pinan-Lucarre1999}%
  \BibitemOpen
  \bibfield  {author} {\bibinfo {author} {\bibfnamefont {J.~P.}\ \bibnamefont
  {Pinan-Lucarre}}, \bibinfo {author} {\bibfnamefont {R.}~\bibnamefont
  {Ouillon}},\ and\ \bibinfo {author} {\bibfnamefont {P.}~\bibnamefont
  {Ranson}},\ }\href {https://doi.org/10.1016/S0009-2614(99)00091-3} {\bibfield
   {journal} {\bibinfo  {journal} {Chem. Phys. Lett.}\ }\textbf {\bibinfo
  {volume} {302}},\ \bibinfo {pages} {164} (\bibinfo {year}
  {1999})}\BibitemShut {NoStop}%
\bibitem [{\citenamefont {Garasevich}\ \emph {et~al.}(1995)\citenamefont
  {Garasevich}, \citenamefont {Slobodyanyuk},\ and\ \citenamefont
  {Yanchuk}}]{Garasevich1995}%
  \BibitemOpen
  \bibfield  {author} {\bibinfo {author} {\bibfnamefont {S.~G.}\ \bibnamefont
  {Garasevich}}, \bibinfo {author} {\bibfnamefont {A.~V.}\ \bibnamefont
  {Slobodyanyuk}},\ and\ \bibinfo {author} {\bibfnamefont {Z.~Z.}\ \bibnamefont
  {Yanchuk}},\ }\href@noop {} {\bibfield  {journal} {\bibinfo  {journal} {Phys.
  Lett. A}\ }\textbf {\bibinfo {volume} {197}},\ \bibinfo {pages} {238}
  (\bibinfo {year} {1995})}\BibitemShut {NoStop}%
\bibitem [{\citenamefont {Grimsditch}\ \emph {et~al.}(1977)\citenamefont
  {Grimsditch}, \citenamefont {Ramdas}, \citenamefont {Rodriguez},\ and\
  \citenamefont {Tekippe}}]{Rodriguez1977}%
  \BibitemOpen
  \bibfield  {author} {\bibinfo {author} {\bibfnamefont {M.~H.}\ \bibnamefont
  {Grimsditch}}, \bibinfo {author} {\bibfnamefont {A.~K.}\ \bibnamefont
  {Ramdas}}, \bibinfo {author} {\bibfnamefont {S.}~\bibnamefont {Rodriguez}},\
  and\ \bibinfo {author} {\bibfnamefont {V.~J.}\ \bibnamefont {Tekippe}},\
  }\href {https://doi.org/10.1103/PhysRevB.15.5869} {\bibfield  {journal}
  {\bibinfo  {journal} {Phys. Rev. B}\ }\textbf {\bibinfo {volume} {15}},\
  \bibinfo {pages} {5869} (\bibinfo {year} {1977})}\BibitemShut {NoStop}%
\bibitem [{\citenamefont {Bizek}\ \emph {et~al.}(1981)\citenamefont {Bizek},
  \citenamefont {Imaino}, \citenamefont {Ramdas},\ and\ \citenamefont
  {Rodriguez}}]{Bizek1981}%
  \BibitemOpen
  \bibfield  {author} {\bibinfo {author} {\bibfnamefont {H.~M.}\ \bibnamefont
  {Bizek}}, \bibinfo {author} {\bibfnamefont {W.}~\bibnamefont {Imaino}},
  \bibinfo {author} {\bibfnamefont {A.~K.}\ \bibnamefont {Ramdas}},\ and\
  \bibinfo {author} {\bibfnamefont {S.}~\bibnamefont {Rodriguez}},\ }\href
  {https://doi.org/10.1002/jrs.1250100119} {\bibfield  {journal} {\bibinfo
  {journal} {J. Raman Spectrosc.}\ }\textbf {\bibinfo {volume} {10}},\ \bibinfo
  {pages} {106} (\bibinfo {year} {1981})}\BibitemShut {NoStop}%
\bibitem [{SM()}]{SM}%
  \BibitemOpen
  \href@noop {} {}\bibinfo {note} {See Supplemental Material at [URL will be
  inserted by publisher] for discussions of the type of detectors and
  resolution of the spectrometers [S1], the circularly polarized Raman spectra
  of Si [S2], and the frequency difference $\Delta\nu$ [S3-S5].}\BibitemShut
  {Stop}%
\bibitem [{\citenamefont {{Margaret M.
  Elcombe}}(1967)}]{MargaretM.Elcombe1967}%
  \BibitemOpen
  \bibfield  {author} {\bibinfo {author} {\bibnamefont {{Margaret M.
  Elcombe}}},\ }\href@noop {} {\bibfield  {journal} {\bibinfo  {journal} {Proc.
  Phys. Soc.}\ }\textbf {\bibinfo {volume} {91}},\ \bibinfo {pages} {947}
  (\bibinfo {year} {1967})}\BibitemShut {NoStop}%
\bibitem [{aB()}]{aB}%
  \BibitemOpen
  \href@noop {} {}\bibinfo {note} {Circular birefringence is a phenomenon in
  which the refractive indices $n_+$ and $n_-$ for right and left circularly
  polarized light, respectively, are different (e.g., for $\alpha$-quartz,
  $n_+=1.54694$ and $n_-=1.54686$.). The refractive indices of light in $(+-)$
  and $(-+)$ are $n_+$ and $n_-$, respectively. The relationship between the
  phonon frequencies and wavenumbers observed in Raman scattering in the
  backscattering geometry is \begin{equation} \setcounter{equation}{1}
  \renewcommand{\theequation}{\roman{equation}}
  \nu_{\mathrm{ph}}=\left(1+\frac{n_{i}}{{n_s}}\right)\nu_{i}-\frac{c}{n_i}q,
  \end{equation} where $\nu_i$, $c$, $n_i$, $n_s$, $\nu_{\mathrm{ph}}$, and $q$
  are the angular frequency of incident light, speed of light, refractive index
  of incident light and scattered light, and phonon frequency and wavenumber,
  respectively. Solving equation (i) and the phonon dispersion relation
  $\nu(q)$ for the sample determines the phonon frequencies observed in Raman
  scattering. Generally, $\nu(q)$ is complicated; however, we can approximate
  it as $\nu(q)=aq+\nu_0$, where $a$ and $\nu_0$ are the slope of the phonon
  dispersion relation and the phonon frequency at the $\Gamma$ point,
  respectively, because the phonons observed in Raman scattering are near the
  $\Gamma$ point. Solving (i) and (ii) for $\nu_\mathrm{ph}$ yields
  \begin{spacing}{1} \begin{equation} \setcounter{equation}{2}
  \renewcommand{\theequation}{\roman{equation}}
  \nu_{\mathrm{ph}}=\frac{1}{1+\frac{c}{n_{i}a}}\left\{\left(1+\frac{n_i}{n_s}\right)\nu_{i}-\frac{c}{n_ia}\nu_{0}\right\}
  \end{equation} \end{spacing} where $\nu_{\mathrm{ph}}=\nu(q)$. From (ii), we
  calculated the observed phonon frequencies $\nu_{(+-)}$ and $\nu_{(-+)}$ in
  the polarization configurations $(+-)$ and $(-+)$, respectively, and
  estimated the order of $\Delta\nu=\nu_{{(+-)}}-\nu_{{(-+)}}$ to be
  approximately $10^{-4}$ cm$^{-1}$, where $\nu_i$ is $10^4$ cm$^{-1}$, $\nu_0$
  is $10^2$ cm$^{-1}$, and $a$ is of the order of $10^2$ cm$^{-1}
  \slash\mathrm{\mathring{A}}$ from the phonon dispersion relation reported in
  previous studies \cite{MargaretM.Elcombe1967}. This is much smaller than the
  observed $\Delta\nu$. Therefore, circular birefringence is ruled out as the
  origin of the observed $\Delta\nu$.}\BibitemShut {Stop}%
\bibitem [{aBC()}]{aBC}%
  \BibitemOpen
  \href@noop {} {}\bibinfo {note} {If Raman experiments are performed in
  scattering configurations other than backscattering, one would expect that
  $\Delta\nu$s are proportional to the wavenumber. However, when circularly
  polarized Raman spectroscopy is performed for phonons propagating in
  directions other than the [0001] direction, both the incident and Raman
  scattered lights become elliptically polarized due to the birefringence of
  the $\alpha$-quartz. In this situation, Raman spectroscopy in the $(+-)$ and
  $(-+)$ configurations cannot selectively observe the Raman peaks of the
  phonons belonging to the upper and lower branches. Conversely, phonons with
  different wavenumbers can possibly be observed by changing the wavelength of
  the excitation laser, even in backscattering. Pine and Dresselhaus~[20]
  reported the linear wavenumber dependence of the E-mode phonon splitting at
  approximately 128 cm$^{-1}$ of $\alpha$-quartz using circularly polarized
  Raman spectroscopy with lasers of different wavelengths. This report suggests
  that the splitting observed in our experiment also exhibits similar
  wavenumber dependence.}\BibitemShut {Stop}%
\bibitem [{\citenamefont {Zhang}\ and\ \citenamefont {Niu}(2015)}]{Zhang2015}%
  \BibitemOpen
  \bibfield  {author} {\bibinfo {author} {\bibfnamefont {L.}~\bibnamefont
  {Zhang}}\ and\ \bibinfo {author} {\bibfnamefont {Q.}~\bibnamefont {Niu}},\
  }\href {https://doi.org/10.1103/PhysRevLett.115.115502} {\bibfield  {journal}
  {\bibinfo  {journal} {Physical Review Letters}\ }\textbf {\bibinfo {volume}
  {115}},\ \bibinfo {pages} {115502} (\bibinfo {year} {2015})},\ \Eprint
  {https://arxiv.org/abs/1502.02573} {arXiv:1502.02573} \BibitemShut {NoStop}%
\bibitem [{\citenamefont {Tatsumi}\ \emph {et~al.}(2018)\citenamefont
  {Tatsumi}, \citenamefont {Kaneko},\ and\ \citenamefont
  {Saito}}]{Tatsumi2018b}%
  \BibitemOpen
  \bibfield  {author} {\bibinfo {author} {\bibfnamefont {Y.}~\bibnamefont
  {Tatsumi}}, \bibinfo {author} {\bibfnamefont {T.}~\bibnamefont {Kaneko}},\
  and\ \bibinfo {author} {\bibfnamefont {R.}~\bibnamefont {Saito}},\ }\href
  {https://doi.org/10.1103/PhysRevB.97.195444} {\bibfield  {journal} {\bibinfo
  {journal} {Phys. Rev. B}\ }\textbf {\bibinfo {volume} {97}},\ \bibinfo
  {pages} {195444} (\bibinfo {year} {2018})}\BibitemShut {NoStop}%
\bibitem [{\citenamefont {Streib}(2021)}]{Streib2021}%
  \BibitemOpen
  \bibfield  {author} {\bibinfo {author} {\bibfnamefont {S.}~\bibnamefont
  {Streib}},\ }\href {https://doi.org/10.1103/PhysRevB.103.L100409} {\bibfield
  {journal} {\bibinfo  {journal} {Physical Review B}\ }\textbf {\bibinfo
  {volume} {103}},\ \bibinfo {pages} {L100409} (\bibinfo {year} {2021})},\
  \Eprint {https://arxiv.org/abs/2010.15616} {arXiv:2010.15616} \BibitemShut
  {NoStop}%
\end{thebibliography}%
\end{document}